\documentclass[]{article}
\usepackage[english]{babel}
\usepackage[utf8]{inputenc}
\usepackage{amsmath}
\usepackage{graphicx}
\usepackage{authblk}
\usepackage{hyperref}
\usepackage{ulem}
\usepackage[margin=1in]{geometry}
\usepackage{natbib}

\title{Roadmap for Reliable Ensemble Forecasting of the Sun-Earth System}
\author[1]{Gelu Nita}
\author[2]{Rafal Angryk}
\author[2]{Berkay Aydin}
\author[2]{Juan Banda}
\author[3]{Tim Bastian}
\author[4]{Tom Berger}
\author[5]{Veronica Bindi}
\author[6]{Laura Boucheron}
\author[1]{Wenda Cao}
\author[7]{Eric Christian}
\author[7]{Georgia de Nolfo}
\author[8]{Edward DeLuca}
\author[9]{Marc DeRosa}
\author[10]{Cooper Downs}
\author[1]{Gregory Fleishman}
\author[11]{Olac Fuentes}
\author[1]{Dale Gary}
\author[12]{Frank Hill}
\author[13]{Todd Hoeksema}
\author[14]{Qiang Hu}
\author[15]{Raluca Ilie}
\author[16]{Jack Ireland}
\author[15]{Farzad Kamalabadi}
\author[8]{Kelly Korreck}
\author[1]{Alexander Kosovichev}
\author[17]{Jessica Lin}
\author[18]{Noe Lugaz}
\author[19]{Anthony Mannucci}
\author[20]{Nagi Mansour}
\author[2]{Petrus Martens}
\author[21]{Leila Mays}
\author[22]{James McAteer}
\author[23]{Scott W. McIntosh}
\author[1]{Vincent Oria}
\author[14]{David Pan}
\author[15]{Marco Panesi}
\author[7]{W. Dean Pesnell}
\author[12]{Alexei Pevtsov}
\author[12]{Valentin Pillet}
\author[24]{Laurel Rachmeler}
\author[25]{Aaron Ridley}
\author[26]{Ludger Scherliess}
\author[25]{Gabor Toth}
\author[27]{Marco Velli}
\author[28]{Stephen White}
\author[17]{Jie Zhang}
\author[25]{Shasha Zou}
\affil[1]{New Jersey Institute of Technology}
\affil[2]{Georgia State University}
\affil[3]{National Radio Astronomy Observatory}
\affil[4]{University of Colorado}
\affil[5]{University of Hawaii}
\affil[6]{New Mexico State University}
\affil[7]{NASA Goddard Space Flight Center}
\affil[8]{Harvard-Smithsonian Center for Astrophysics}
\affil[9]{Lockheed Martin Solar and Astrophysics Lab}
\affil[10]{Predictive Science Inc.}
\affil[11]{University of Texas at El Paso}
\affil[12]{National Solar Observatory}
\affil[13]{Wilcox Solar Observatory}
\affil[14]{University of Alabama in Huntsville}
\affil[15]{University of Illinois at Urbana Champaign}
\affil[16]{ADNET Systems, Inc.}
\affil[17]{George Mason University}
\affil[18]{University of New Hampshire}
\affil[19]{California Institute of Technology, Jet Propulsion Laboratory}
\affil[20]{NASA Ames Research Center}
\affil[21]{Community Coordinated Modeling Center}
\affil[22]{New Mexico State University, Sunspot Observatory}
\affil[23]{High Altitude Observatory}
\affil[24]{NASA Marshall Space Flight Center}
\affil[25]{University of Michigan}
\affil[26]{Utah State University}
\affil[27]{University of California, Los Angeles, Earth Planetary and Space Sciences}
\affil[28]{Air Force Research Laboratory}

\date{\today}
\begin{document}
\bibliographystyle{unsrtnat}
\maketitle

\begin{abstract}
The authors of this report met on 28-30 March 2018 at the {\it New Jersey Institute of Technology}, Newark, New Jersey, for a 3-day workshop that brought together a group of data providers, expert modelers, and computer and data scientists, in the solar discipline. Their objective was to identify challenges in the path towards building an effective framework to achieve transformative advances in the understanding and forecasting of the Sun-Earth system from the upper convection zone of the Sun to the Earth’s magnetosphere. The workshop aimed to develop a research roadmap that targets the scientific challenge of coupling observations and modeling with emerging data-science research to extract knowledge from the large volumes of data (observed and simulated) while stimulating computer science with new research applications. The desire among the attendees was to promote future trans-disciplinary collaborations, and identify areas of convergence across disciplines. The workshop combined a set of plenary sessions featuring invited introductory talks and workshop progress reports, interleaved with a set of breakout sessions focused on specific topics of interest. Each breakout group generated short documents, listing the challenges identified during their discussions in addition to possible ways of attacking them collectively. These documents were combined into this report \-- wherein a list of prioritized activities have been collated, shared and endorsed.
\end{abstract}









\section{Workshop Organization}

This workshop, entitled {\it Roadmap for Reliable Ensemble Forecasting of the Sun-Earth System}, was supported with funds from the National Science Foundation. The workshop organizing committee comprised Drs. Gelu Nita (PI, NJIT), Dale Gary (Co-PI, NJIT), Rafal Angryk (Co-PI, Georgia State University), Farzad Kamalabadi (University of Illinois at Urbana-Champaign), and Scott McIntosh (High Altitude Observatory). The workshop was attended by 47 active participants representing 25 distinct US institutions. In addition, Drs Michael Wiltberger (Section Head, GEO/AGS), Ilia Roussev (Solar and Heliospheric Physics Program Director, GEO/AGS) and David Boboltz (Solar and Planetary Research Program Director, MPS/AST) participated as observers.

The first day of the workshop featured nine invited introductory presentations, after which the workshop participants split into two breakout groups, \textit{Data Providers/Analyzers} and \textit{Modelers}. The participating computer and data scientists joined one or another breakout group, based on their expertise and preference. The charges to the breakout session participants were to i) \textit{Identify current challenges and potential directions for progress toward advancing predictive capability} and ii) \textit{Identify research directions within each group that can help accelerate progress in the other groups in light of the workshop objective}. The breakout session was followed by a brief plenary session in which progress reports were presented to the collected participants.


On the second day, the workshop participants split into three breakout groups, namely \textit{Data Providers}, \textit{Modelers}, and \textit{Computer and Data Scientists} to continue their discussions, in light of the previous day reports. Following the previous day's format, the breakout sessions were interleaved with plenary sessions for reporting progress.


The last day of the workshop was dedicated to the groups presenting their findings and documenting their sessions. The collected participants had the opportunity to discuss the findings and comment on any of the topics presented. Following the workshop, the Organizing Committee members edited the breakout group reports and combined them into this document.

This document was distributed in draft form, verified, and then endorsed by the workshop participants as an accurate reflection of their efforts over the three days of activity.

\section{Emerging Research Directions in Sun-Earth System Forecasting}
Progress toward reliable forecasting of the Sun-Earth system requires concurrent advances in computational modeling and data science, as well as their harmonious integration. Embedded in all computational modeling efforts — whether first-principles (FP) or empirical — is a framework based on observational data. The distinction between FP and empirical models is not sharp. All FP models used in space weather modeling have some empirical aspects. For example, the interesting physics in magnetized plasmas take place in current sheets that are simply parametrized in magneto-hydrodynamics (MHD), while a flux rope based CME model may use an empirical relationship between the CME speed and the energy of the flux rope. Global-Sun flux-transport models that are used to predict the solar wind and the solar cycle are based on photospheric magnetograms as well as on information from helioseismology about the meridional circulation and far-side imaging. Models of particle acceleration and transport are constrained by vector magnetograms and multi-wavelength spectro-polarimetric data covering the whole range of electromagnetic spectrum, from radio waves to gamma rays. The magnetic field measurements that are at the core of space weather models are based on FP modeling of the radiative magnetohydrodynamics of the photosphere and chromosphere and inversion of radiative transfer models. Therefore, for comprehensive understanding of the underlying physical processes that govern space weather, and in longer term, the space climate, it is essential to develop a complex program that provides synergy of data and modeling for all scales, from the global dynamo to magnetic energy release in the form of CMEs and energetic particles.

In general, FP models solve a complicated system of equations based on some physical approximation of the system and they tend to provide more comprehensive information than empirical models. On the other hand, empirical models typically run faster than FP models, which may make them more suitable for ensemble forecasting. Empirical models are unlikely to perform well for extreme events that are outside the range of the events that was used to build them. FP models can in principle provide reasonable forecasts for extreme events assuming that the physical and numerical approximations remain valid. FP models can also provide insight into the physical processes, while empirical models provide results without explicit physical meaning, although such physical meaning may be inferred from the results.

All computational models used for forecasting the sun-Earth system rely on observational data to guide or constrain their forecasts.  As such, the issues of making the relevant data available, robust, and appropriately processed are essential in advancing the state of forecasting.  Sophisticated processing of the relevant data for the task of inference from observations may be approached through inversion techniques and machine learning, and systematic integration of FP models and data may be approached through data assimilation.  Machine learning enables training an explicit or implicit model, learning its attributes, and inferring physical parameters from the model.  When the models are explicit and parametric, machine learning can be regarded as a particular form of empirical modeling. In general, empirical models require careful selection and preprocessing of input data (a training set) and this typically requires physical insight and potentially the application of some FP methods.  Data assimilation requires the characterization of data quality (e.g., uncertainty quantification) as well as data robustness, often requiring careful processing of observational data for incorporation into FP models.

We expect that the most successful space weather models will optimally combine empirical and FP models, and incorporate data assimilation and machine learning techniques. As such, advancing research frontiers in these directions is essential for meaningful progress. For example, an empirical model based on machine learning may be optimal to predict whether a certain active region will produce a CME of a certain speed range and a certain time interval. This information may be used to feed an ensemble of FP or empirical CME models that follow the evolution of the CME from the Sun to the Earth. These simulations will be strongly affected by the background solar wind model, which is likely to be FP based.  Interaction of multiple CMEs or extreme CMEs will probably require FP models, but single non-extreme events may be well predicted with empirical models. Data assimilation may be used to find the best members of the ensemble. The output of the CME models can be used to drive FP-based magnetosphere models that can predict localized space weather effects on the ground. The flare and CME prediction could also be used to drive FP solar energetic particle models.

The above example is just one possibility, and there are many alternative approaches that would require systematic investigation, but it shows the intricate interplay between observational data and first-principles models. It also highlights that the empirical models should provide a range of predictions rather than single output. Similarly, FP models need to be run in ensemble mode to provide a probabilistic forecast.

In summary, optimal integration of data and FP models is a non-trivial problem that will require research across multiple disciplines, experimentation, innovation, and systematic evaluation.

\section{Consensus}
The workshop participants drafted a common purpose, a banner to march under:

{\it Improve Understanding of the the Sun-Earth System to Advance Predictability Through a Systematic Synthesis of Observation, Modeling, and Data Analytics.}

The following is a list of prioritized consensus recommendations, first listed by group in sections 3.1--3.3 and then combined into a cross-cutting set of recommendations in section 3.4.

\subsection{Modelers Group}
\begin{enumerate}
\item Support necessary research to utilize first-principles computational models for physical understanding and forecasting by developing appropriate computational algorithms and resources to enable efficient high-fidelity simulations with predictive capability.
\item Ensure the availability and optimal use of relevant data needed for the development of data-driven and data-assimilation models, facilitate benchmarking and inter-comparison of models, and create funding mechanisms where multi-disciplinary teams with expertise in solar/heliospheric physics, data science, and applied mathematics/statistics are encouraged to work together.
\item Address the research challenges that hinder the seamless integration of models and data. Support research necessary to develop models that incorporate a variety of time-dependent constraints, and  cross-disciplinary research into data-assimilation techniques for the temporally and spatially sparse datasets that exist in the solar/heliospheric environment.
\item Support research toward developing algorithms and resources that enable the use of ensemble simulations and related quantification of  uncertainties in predictions.
\item Develop and adopt validation methods and benchmark data for meaningful quantification of forecasting skills.
\item Investigate the role of first-principles physical model simulations in generating training data for machine learning techniques.
\end{enumerate}

\subsection{Data Providers Group}
\begin{enumerate}
\item The currently available suite of observations are not adequate to significantly advance the forecasting of nascent and eruptive space weather phenomena. Identify the most glaring data gaps and the most essential missing observations necessary to fill the data gaps. Explore mechanisms for generating the needed observations and observational data products.
\item Standardization of metadata, and data quality measures are imperative if data are to be implemented in a data-driven, or assimilative, forecast methodology.
\item We must develop a repository of space weather ``events'' (pulled from previous case studies from SHINE, AGU, LWS, etc.), including the nascent conditions around them, that can provide a broad range of benchmarks for forecast method development. The methodology of the SDO/HMI ``SHARPS'' could be exploited and adapted, but only exists beyond 2009 and so work is required for events in earlier solar cycles.
\item We must advance/develop a community ``observable synthesis'' framework that can reasonably be applied in forward, inverse and assimilative methodologies to permit ``apples-to-apples'' comparison between observations and models.
\item Borrowing terminology from the meteorological community, a framework of Observing System Experiments (OSE) and Observing System Simulation Experiments (OSSE) should be considered for space weather to critically assess the impact of present and future data/measurables on the forecast accuracy and skill of the models. The aforementioned point is essential in this process.
\item As an example of future need, routine measurement of the global chromospheric and coronal plasma environment is desired by many, especially the (vector) magnetic fields. This research activity explicitly requires the funding and development of advanced inversion techniques and education that goes with them \-- such methods should not be ``black boxes'', requiring cross-disciplinary research with contributions from the relevant applied mathematics/statistics communities.
\end{enumerate}

\subsection{Computer Science Group}
\begin{enumerate}
\item Create data set benchmarks, to enable comparative studies of machine learning (ML) models.
\item Take advantage of state-of-the-art machine learning/data science models (e.g., deep learning).
\item Develop scalable and distributed machine learning algorithms for solar Big Data mining.
\item Address challenges with employing machine learning models in operational settings (e.g. real-time data availability, adjustment of ML models as the solar cycle evolves.
\item Develop event-guided image data compression techniques for easier access to Big Data. Establish centers for holding and processing the solar Big Data. Evaluate alternative high performance computing and storage solutions: HPC vs. Cloud Computing vs. Computing Clusters vs. GPUs.
\item Support knowledge dissemination and training of solar data scientists
\end{enumerate}

\subsection{Combined Consensus Topics}
There are common themes among the breakout groups that should be highlighted as important agreed-upon consensus items:
\begin{enumerate}
\item Development of \textbf{new and more capable modeling approaches} that integrate data and first-principles models was highlighted in items 3, 4, and 3 of the modeler, data provider, and CS groups, respectively. This development requires research in computational algorithms for data assimilation, machine learning, and advanced inversion techniques.
\item \textbf{Quantifying uncertainties} in data and model forecasts was highlighted in items 5 and 2 of the modeler and data provider groups, respectively.
\item \textbf{Benchmarking} certain datasets for use by modelers, which are self-contained online repositories of all relevant data and derived data products for individual events or time periods. This was highlighted in items 2, 3, and 1 of the modeler, data provider, and CS groups, respectively.

\item \textbf{Cross-disciplinary research and funding mechanisms} involving investigators in solar/heliospheric physics, data/computer science, and applied mathematics/statistics was highlighted in items 2, 6, and 6 of the modeler, data provider, and CS groups, respectively.
\item Increased support for the \textbf{creation of infrastructures} to create the needed data, algorithms, and computational capabilities was highlighted in items 1, 3, and 6 of the modeler, data provider, and CS groups, respectively.

\end{enumerate}
The remainder of this report provides the detailed motivations that led to these priorities, as they emerged from the breakout-group discussions.

\section{Modelers Group Breakout Session Report}
The modeling breakout group, comprising researchers from the solar-terrestrial research community, deliberated on identifying challenges/gaps and emerging research directions needed to advance the computational modeling component of Sun-Earth system forecasting.  The following is a summary of the findings and recommendations endorsed by the modeling breakout group.









\subsection{Address the main challenges and bottlenecks in utilizing first-principles computational models for forecasting}

\subsubsection{Incomplete physics in first-principles models}
Improving forecasts with more complete physics requires research in computational physics, computational mathematics, and scientific computing.

The solar corona is a magnetized plasma which transitions through different regimes in different regions in space, from the chromosphere (high beta) to the transition region and corona (low beta) into the solar wind (beta order one). At the same time, the range of validity over different spatial scales of the plasma model used (typically Compressible Magnetohydrodynamics) changes depending on location. This impacts the physics that may be captured by the model, specifically via thermodynamics and instabilities that depend explicitly on transport and dissipative coefficients such as thermal conductivity, resistivity and viscosities. Time-scales are also challenging: the characteristic time-scales of the Poynting flux into the corona (~minutes) and the time-scale of large scale evolution as well as solar wind transport to 1 AU (days) are widely separated. Therefore, even for the background pre-eruptive coronal state, a model of the wave and turbulence flux and its dissipation are required to provide realism. The time-dependent forcing of the model then needs adjustment on timescales compatible with the full 3D model as well.

With a given coronal background, initializing a CME \textit{ab initio} would require the complete history of the photospheric magnetic flux as well as a physically realistic description of the relaxation processes (magnetic reconnection, magnetic helicity inverse cascade) leading to the accumulation of magnetic energy in the solar corona. In reality, it is at present impossible to develop the initial configuration and see it become unstable self-consistently. Therefore, it is customary to use the highest resolution magnetograms available (scalar and vector) as boundary conditions together with the so-called flux-rope insertion model. A minimal requirement for flux-rope insertion is to ensure that the chirality and axial magnetic field of the flux rope is correct, i.e. consistent with the filament actually seen to erupt on the Sun. An important step forward would be to provide for such synoptic data, and have an automated detection and parametrization of such properties of solar filaments.

Hence, from the research point of view it is important: A) to make progress on a self-consistent global 3D solar wind model, with sub-grid models for the physics at the scale where dissipative effects are fundamental (i.e. models for the kinetic physics in reconnection layers). Better models for the evolution of the Poynting–flux fluctuations (i.e. turbulence models for open and closed magnetic fields) are required to obtain correct parametrizations in global models. B) to make progress on ab-initio evolution of the coronal fields subject to footpoint changes dictated from photospheric magnetograms (i.e. the driving electric field) to self-consistently obtain the initial magnetic field configuration which then erupts into the corona.

\subsubsection{Computational resources}
Both the incorporation of more complete physics and ensemble runs increase computational cost, therefore requiring research on computationally efficient cyberinfrastructure (algorithms and platforms).

A primary limitation in enabling high fidelity simulations for forecasting purposes is the need for large computational resources. Ensemble forecasting requires conducting multiple such simulations, which probe large parameter spaces in order to provide a range of possible prediction/states of the system and hence require substantial computational resources.  Another limitation pertains to the physical separation of observational data and modeling platforms; often, observational data needed by both FP and empirical models is stored on platforms that are physically separated than those where models are run. Therefore, access to data is cumbersome and adds additional delays and strains for real-time forecasting efforts.

Algorithm development is rarely integrated into the funded efforts pertaining to the development of space weather prediction models. Partnership with research communities that have significant algorithm development expertise have often been underutilized in solar-terrestrial modeling.  Yet, such communities are supported by several programs within CISE and MPS and can offer rich and fruitful partnerships with AGS-STR, and AGS-SW.

\subsection{Ensure the availability and optimal use of relevant data, and characterization of forecasting uncertainty}

Another major challenge in utilizing first-principles computational models for forecasting lies in the availability and coverage of the specific data products needed for time-dependent driving and/or assimilation by models. Although there is a variety of useful science data products routinely made available by the heliophysics system observatory, they form a heterogeneous data set that is not natively ready for insertion into models. One key example is the use of photospheric measurements of the magnetic field as boundary conditions for coronal models (empirical, magnetofrictional, MHD). In this case, the global structure of the corona is fundamentally influenced by the polar and far-side magnetic fields, both of which are not currently observed at all, and not well described by flux-transport models. This places strong limits on the accuracy of data-driven and data-constrained coronal models. Therefore, it is of high priority to develop  robust flux-transport and, in perspective, more complete dynamo models, which assimilate flux-emergence and far-side helioseismology data. Furthermore, for the routine magnetic data products that are obtained, such as full-disk line-of-sight and vector magnetic fields measured by HMI, it is an active area of research to convert this information into a product that can self consistently drive the boundary of MHD models (i.e., a combination of both time-dependent electric and magnetic fields, as well as plasma flows). Processing such data has traditionally required multi-year research efforts by teams with observational and modeling expertise, and it is crucial to leverage such efforts into building community-wide derived data-products that have a solid physical basis and large temporal coverage. In this manner the community can accelerate the process of developing and testing data-driven and FP models, and facilitate the benchmarking and inter-comparison of such models for a large number of cases, where a `standard’ set of driving and benchmark data products is crucial for building consensus.

Similarly, magnetograms are often used to generate a single potential or non-linear force-free field (NLFFF) model that is used to drive other models.  Since the measurements have uncertainties, this should be taken into account when generating a magnetic field model, such that a wide variety of magnetic field models could be generated, enabling ensemble simulations to be conducted.

Another example includes L1 measurements. These measurements are often used to drive a wide variety of geospace models in order to predict the near-Earth space weather.  There are multiple problems with this simplistic approach: Because the satellite(s) that measure the solar wind and magnetic field are typically ~30 $R_{\rm E}$ off the sun-Earth line, the measurements may not accurately represent the true solar conditions that encounter the nose of the magnetosphere. Also because the satellites are off the Sun-Earth line, the exact timing of when the solar wind will encounter the magnetosphere is often unclear and could be off by several minutes. Global magnetospheric models typically assume that the solar wind and IMF enter the domain as a flat plane, which may or may not be correct.  Further, they often assume that the plane normal is oriented in the GSM-X coordinate, which is definitely not correct almost all of the time.  This causes inaccuracies in both timing and conditions that encounter the Earth, since physical constraints such as the divergence of the magnetic field must be taken into account when non-tilted planes are assumed. Research needs to be conducted in order to determine how the non-ideal and incomplete L1 measurements affect driving conditions and the uncertainty in the model predictions.

More broadly, the impact of data limitations, spatio-temporal sampling, and measurement uncertainties in data that are incorporated into first-principles models must be characterized and quantified.  Such endeavors require facilitating the creation of environments and funding mechanisms where teams with disparate expertise (computational mathematics, applied statistics, and plasma physics) are brought together to bridge the knowledge gaps and tackle such multi-disciplinary pursuits in a cohesive manner.

\subsection{Address the research challenges and gaps that hinder the seamless integration of models and data}

\subsubsection{Time-dependent constraints}
Current developments in first-principles modeling do not generally take full advantage of time-dependent constraints, including boundary conditions. Time-dependent data are primarily used for data-model comparisons rather than systematic integration of the data into the models themselves. This is especially true in the solar-heliospheric community (vector magnetograms, “ingesting” non-linear force free models, CME observations, etc.) but also in the magnetosphere. However, progress in forecasting requires the development of models with the capability to systematically incorporate a large variety of time-dependent constraints. Such developments can enable models that are versatile enough that would allow investigating how different data impact the physical specification and understanding, and forecasting performances.

\subsubsection{Data assimilation research}
For the past decades, data assimilation techniques have been successfully developed and employed for specification and forecasting in meteorology and oceanography. More recently these methods have also been successfully used for the near-Earth space environment and it is expected that data assimilation will also provide a powerful path toward forecasting for the solar and heliospheric environment. However, the development of data assimilation models will require several important and necessary research investigations and advances that demand cross-disciplinary efforts.

The foremost requirement for the advent of these models is the need for research into and development of appropriate data assimilation techniques that are adequate for the temporally and spatially sparse datasets that exist in the solar/heliospheric environment. Although a variety of different techniques have been developed over the past decades in other domains, whether these techniques are adequate for use with current solar/heliospheric models needs to be investigated. Furthermore, it is important to understand that sensitivities to initial conditions and drivers in different domains might require different approaches; a simple application of proven techniques that work in meteorology or oceanography might not work for the solar/heliospheric environment.

The implementation of data assimilation techniques into solar/heliospheric models is furthermore expected to require a significant effort and new model developments might be required (i.e. 4-d Var requires the adjoint of the physical model). In case of the implementation of ensemble techniques that have recently gained much prominence in data assimilation, methods that address the important issue of the generation of proper ensembles need to be researched and developed.

One of the prime computational bottlenecks of modern data assimilation techniques is the computation of the temporal evolution of the model error covariances. These covariances are critical for the data assimilation models and effective ways to calculate them for the solar/heliospheric environment need to be researched and developed. One prominent method for this is the use of ensemble (particle) filtering Ensemble Kalman filtering (EnKF) and requires research on the appropriate online generation of the ensembles and approximation of the covariances.

Furthermore, observations that represent sub-grid scale processes, or unmodeled physical processes, must be processed in some way before they can be assimilated. Otherwise, the inherent inconsistency between the model output and the assimilated observations cannot be resolved via assimilation, yielding unexpected results and degraded forecasts. It is important to identify when the observations have a significant component that is incompatible with the first-principles modeling and to develop a processing strategy for the data before it is assimilated. Machine learning methods could be beneficial to both classify the observations where this is an issue, and possibly to develop an automated statistical technique to pre-process the observations before assimilation.

Finally, since observation noise is inevitable, an important research question is: what is the impact of data noise on the data assimilation? In every assimilation scenario, criteria should be available for assessing the acceptable level of noise. It is an open research question both how to assess the acceptable noise level before assimilating data into models, and how to process the observations to reduce noise without introducing biases. In the case of Gaussian noise and linear models, Kalman filtering (or Bayesian filtering more generally), observational noise is accommodated as part of the formalism and is well understood. For the FP models, which are intricately non-linear, the Ensemble Kalman Filtering is a powerful, but computationally demanding approach. It is an important and cross-cutting topic to address optimal ways of assimilating noisy data. The possibility of biased observations should be included as well.

In summary, appropriate techniques need to be researched and developed to:

\begin{itemize}
\item Develop sparse data assimilation (in space and time)
\item Address the issue of model biases during the initial stages of data assimilation
\item Develop ensemble simulations to obtain covariances for model variables and observations for 4D Var data-assimilation models, as well as the Ensemble Kalman Filtering method for non-linear FP space weather models.
\item Develop robust capability for incorporating sub-grid scale models or observations that are sensitive to unmodeled physics
\end{itemize}

\subsection{Investigate the algorithmic challenges in ensemble runs, aggregation of results, and quantification of uncertainties}

While prediction of different states allows operators to make decisions, it is important to understand the uncertainty in the prediction. Uncertainties allow operators to make better decisions, since they can then understand the trustworthiness, or robustness, of the results. In order to determine the uncertainty in the prediction, the models need to be able to provide those uncertainties in some form.  There are multiple approaches for accomplishing this:

\begin{itemize}
\item Statistical uncertainties in the model results can be generated for a wide variety of different conditions before the predictions are made.  This can be done by varying parameterizations and drivers in the model to explore how the uncertainties depend on these. Then, when a prediction is made, these statistical uncertainties can be applied.
\item An ensemble of predictions can be made, where the ensemble contains members that span the uncertainties in the parameterizations, drivers, and boundary conditions. This way the prediction uncertainties can be specified for the particular event.
\end{itemize}

Both methods have the possibility of taking a great deal of computational resources, since in order to accurately determine the uncertainty, the number of simulations could be very large. At this time, there is not a clear understanding of how to accurately specify the model uncertainty using either technique. Most FP models have a large variety of model parameterizations, each of which can cause uncertainty in the model results.  Model boundary conditions and/or drivers can be 1D, 2D, 3D, or even 4D, which means that a large number of ensemble members are needed in order to capture how the uncertainty in these conditions drive uncertainty in the results.  It is unclear how to optimize this problem; it is a research topic that requires participation of investigators from multiple discipline, in particular applied mathematics and statistics.  For example, by precomputing the statistical uncertainties, the real-time runs can be much faster, but the uncertainties may not be as accurate.  Conversely, in order to capture the breadth of uncertainties in the model, the number of ensemble members that needs to be run in real-time could be computationally prohibitive. Because of this, incorporating the uncertainties in drivers and parameterizations is often simply ignored, and single simulations are conducted for prediction, providing no error estimation at all.

However, in order to actually run ensembles of FP models, it is important to understand the uncertainties in the drivers, boundary conditions, initial conditions, and model parameterizations.  Often these drivers and/or boundary conditions are specified by either other models or measurements. It is then important for those models or measurements to provide both their state and uncertainty, so that ensemble models can perturb these states to drive their models.

Further, because the space science community has relatively modest experience with ensemble simulations and data assimilation, there is little collective experience in how to interpret both the results and the uncertainties in the results. Selection of appropriate ensembles requires approaches to searching the parameter space in some optimal and computationally tractable manner, and concurrent estimation of uncertainties that would be needed for systematical aggregation of ensemble simulations.  These are research topics that have traditionally been pursued under the domain of statistical sampling research.

The research that is needed to tackle these open questions is not domain specific, meaning that the technical and algorithmic challenges that the solar and heliospheric science faces in order to implement ensemble forecasting and uncertainty quantification may be similar to the challenges that other communities face in their pursuit of data assimilation and ensemble forecasting.  Therefore, meaningful progress toward these goals would necessitate interdisciplinary teams encompassing space physics, computational mathematics, statistics, decision theory, and algorithm design.

\subsection{Develop and adopt methods for meaningful quantification of forecasting skill}

Science and forecasting validations are both critical for advancing the science in space weather models and for assessing the state of our current forecasting capabilities. The space physics community should move beyond case studies. We should be running and validating more of our FP models in forecasting mode in order to quantitatively track progress and understand the influence of model developments.  All data from validation studies should be made available upon publication for reproducibility.

Model validation and their statistical methods is a research field in its own right.  Quantifying performance in space weather research can greatly benefit from cross-disciplinary collaboration between the terrestrial weather verification, space weather, statistics, and computer science communities.   We can leverage statistical approaches for model verification from the weather community by working directly with these experts to find analogous verification scenarios, and possibly directly using their verification algorithms once they have been adapted to space weather datasets. Through collaboration with the statistical signal processing/computer science communities, we could reduce the human bias by developing classification algorithms to select a representative sample of benchmark events and additionally select the most appropriate events relevant to the science or forecasting question. To quantify the forecasting skill of models, a statistically significant set of events/epochs must be determined. Such a set must include all representative scenarios (no/weak/strong/extreme events). Once agreed upon, this set of benchmark events/epochs (community/question-dependent) should be used by all. This list can be updated. Obviously, training and verification sets should be kept separate.

The choice and adoption of metrics (such as skill scores) need to be settled in the near term, ideally within one year or less. One proposition is to have a set of parameters, thresholds, time durations, and forecast window for each domain (e.g. flare class, total flare energy, CME speed, SEP flux, IMF Bz, magnetopause location, Dst value, Delta B magnitude, etc.).  The iCCMC-LWS workshop in April 2017 was a starting point for this. To maintain momentum, a small-scale interdisciplinary working meeting for all of these domains may be beneficial.

In summary, to meaningfully compare the performance of difference models it is important to (1) standardize model inputs/settings for each model type (to the extent possible) (2) have access to processed data (with uncertainties) for model inputs and for validation (3) have metadata for all aspects (data inputs, model settings/inputs, model outputs, data used for comparison) (4) use a statistically significant set of events/epochs (5) keep training and validation sets separate. Note that there is still some value in validating models running in "forecast mode" using non-standardized inputs.

\subsection{Investigate the role of first-principles physical model simulations in generating training data for machine learning techniques}

In general, first-principles (FP) model simulations can be used to generate training data for machine learning (ML)

\begin{itemize}
\item If the fidelity of the FP model is sufficiently high that it reproduces reality without major errors and biases and
\item if running the FP model can produce training data more efficiently than observations.
\end{itemize}

There may be opportunities to use ML to emulate FP models to speed up the run time. This may work well if the relationship between input and output of the FP is not very complex and the training can be done efficiently. Another approach would be to use FP models to enhance the training set of the ML with cases that do not exist in the available observations, for example extreme events. This, of course, requires that the FP model is reliable enough for these extreme/exceptional cases. Finally, the FP may be used to train the ML to provide output where observations are not available. For example, the FP model can produce magnetic perturbations on the whole surface of the Sun even where there are no magnetogram data. Or it can provide output like location of new emerging regions, which were not observed on the Earth side. The well trained empirical model may provide output that is of comparable quality as the more expensive FP model.

\section{Data Group Breakout Session Report}

Over the course of the workshop the data (providers) group was challenged to discuss and answer a set of questions to highlight the challenges and hurdles that must be overcome to reach a state where “Reliable Ensemble Forecasting of the Sun-Earth System” was a possibility. In discussing these points a set of consensus opinions was developed - the synthesis of which can be tracked in the notes of the discussions compiled by the rapporteur.

The data providers group chose to respond sequentially to the questions set by the organizing committee. The following narrative is built out of the discussion elements they had around the utilization of data and diagnostics to advance the goals of the workshop.

\subsection{Identify current challenges and potential directions for progress toward advancing the state of prediction--solar magnetic eruptions (emerging magnetic flux, flares, energetic particles, and CMEs}

\begin{itemize}
\item \textbf{What are the main challenges and bottlenecks in utilizing available multi-modal data for forecasting?}

The space weather enterprise would benefit by considering parceling the elements of the space weather enterprise and performing a set of OSE/OSSE exercises to assess the impact of individual observations/data on the modeling activities. Such exercises are critical to assessing and developing assimilative meteorological forecast models. Driving better models using observationally-driven solar wind models; incorporation of photospheric, chromospheric and coronal magnetic fields, plasma flows, flux-transport global-Sun models, etc.

\item\textbf{What are the challenges in robust and routine acquisition and processing of relevant data?}

There is general consensus that the currently available, and archived observational data, are limited in their ability to advance physical understanding to a point where they will drive forecast skill upward. It was also acknowledged that this was not the point of the workshop. Of the pertinent issues in this area would be the standardization of data products available for incorporation into forecast models. For example, ground-based observations tend to suffer measurably more from noise than their space-based equivalents.

\item\textbf{If data uniformity is important, what are the challenges of obtaining support for such synoptic data?}

The group acknowledges that there are considerable challenges to maintaining a synoptic data program. There is an ongoing Operation and Data Analysis cost associated with the acquisition of “routine” synoptic data - often this data is not widely considered as scientifically cutting edge and becomes difficult to justify in a tight funding environment. Typically then such programs are reduced to endowed observatories and/or National Centers where the funding issue also eventually arrives.

\item\textbf{What are the computational-power driven and/or storage-driven bottlenecks of empirical modeling for real-time forecasting?}

While storing archival data is not necessarily an issue as very little of what has been acquired historically has been digitized and made discoverable - an acknowledged issue - those libraries require little physical space. There is considerable effort required to take the decades of solar, in-situ and geospace measurements that live in various archives and make them discoverable for historical forecast/climatology experiments that are essential for modeling long-term evolution of the solar variability. With more advanced data products, (e.g., spectral imaging, imaging spectropolarimeter, radio-spectroheliograms, etc that are multidimensional data cubes) at high cadence will present a storage challenge and hence impact budgets. Furthermore, strategies will need to be developed on what level of data must be maintained and stored in perpetuity. Implementing those next-gen data into a data-driven or assimilative models will require good standardization, assessment of data quality and good documentation of methods  those methods in terrestrial meteorology are particularly storage-heavy, particularly when considering ensemble models.

\item\textbf{What are the challenges in quantifying data quality (e.g., calibration, uncertainty quantification, etc)?}

Beyond some of the standardization mentioned above, data providers expect users to know and abide by limitations stated by them, although the users may not know or want to do so. The data providers do need to provide tools, such as spectral synthesis methods, that allow people to use the data effectively for other purposes, including forward modeling the observations for one-to-one comparison between observation and model, such methods are also a necessary ingredient for assimilative frameworks.


Versioning and frozen benchmarks are needed when processing data, so that data from different eras can be compared. Versioning of data-prep algorithms and associated data would aid in comparing higher level data products made at different times. Data citation is an issue, with the advent of Digital Object Identifiers (DOIs) there may be methods to qualify large data sets, especially those that have gone through many different calibration pipelines.

\item\textbf{What are the algorithmic developments that are needed for ingestion into first-principles models?  What is the role of data assimilation?}


Again, standardization is critical. Modelers need to communicate the standards and error tolerances for the data to be used as boundary conditions in their models. Standardization of metadata, especially of ground-based data will help. If a multi-scale modeling approach is to be undertaken, then numerical methods that rapidly and accurately aggregate and average the observational data to the required/prescribed spatial and temporal scales by the simulation are necessary.

\item\textbf{What is the potential of incorporating additional (new) data sources for forecasting?}

Our community needs to incorporate the OSE/OSSE approach to space weather forecasting. In such frameworks new observations/diagnostics can be inserted and assessed for “impact” on forecast skill, etc.

\item\textbf{What is the role of emerging advances in data science such as machine learning and feature recognition in forecasting?}

It is widely believed that machine learning and computer vision techniques will be of use in the forecasting of space weather events and baseline states of the inner heliosphere. It is note that only the last twenty years of observational data is in a format, is readily available, and documented that it could be used to explore the individual events and/or the climatology of the system. The latter experiments would appear to be easily amenable to advanced CS techniques, but so little of our historically archived synoptic data is in a condition that it can be considered discoverable.

\item\textbf{What future research directions have the highest potential for accelerated progress toward improved forecasting by utilizing data in unprecedented ways?}

There was no consensus on this answer although options discussed included the exploitation of whole solar atmosphere observations to extend longitudinal evolution of the Sun at different epochs of the solar cycle that are informative (e.g., looking at time histories of recurrent coronal holes, active regions in terms of flare and CME productivity, etc.). We have seldom tried to understand the longitudinal behavior with advanced CS algorithms that can, and should, be compared to human analyses for skill.


\item\textbf{What are main challenges in the integration of multiple space data sources?}

Often issues lies in the level of expertise and familiarity of the user with data. Many tools and manipulation methods exist to integrate, coalign, regrid, data sources onto common scales for ease of use.

\item\textbf{What are main challenges and bottlenecks in allowing computer scientists to provide their expertise in supervised machine learning from space data?}
The classical approach is that solar physicists propose a question and use data and modeling to answer that question. Machine learning (and data mining) in particular must be tailored to allow for all of the previous work in classification and understanding. Feedback mechanisms are therefore required between provider, user and computer scientist implementor. Concerns are high that forecast skill requires physical understanding, but that the machine codes could highlight areas of correlation to pursue for physical causation. An item of some intersection here would be the subsetting of data, of a list of flares, which had CMEs, SEPs, etc., and many many more flavors such that patterns can be sought. See also the above discussion on climatology and longitudinal investigations linking active regions and solar features over rotations.
\end{itemize}

\subsection{Identify research directions within each group that can help to accelerate progress in the other group in light of the workshop objectives}

\begin{itemize}
\item\textbf{What are the current and upcoming data sources that can be incorporated into first-principles models?}

A very broad range of synoptic-scale measurements are available in radio and optical wavelengths: providing line-of-sight and vector magnetograms of the photosphere, crude line-of-sight magnetograms of the chromosphere and linear polarization measurements of the coronal magnetic field. In addition, to these helioseismic and dynamical (wave) data are available to probe wave/magnetism relationships. The Mauna Loa Solar Observatory provides measurements of the K-Corona that are complemented by spectropolarimetric measurements of coronal emission. The Expanded Owens Valley Solar Array can provide maps of chromospheric and coronal magnetism at contours of constant temperature. The Dunn Solar Telescope can provide synoptic magnetic field information on filaments and coronal holes. The Wilcox observatory continues to be critical in providing polar field measurements.

In the next few years we’ll have in situ measurements from Parker Solar Probe (PSP) and Solar Orbiter (SolO) to complement existing space-based capabilities (SoHO, SDO, Hinode, etc) - SolO will also have remote sensing instrumentation and be able to sample high solar latitudes. At the same time, the 4-m aperture of the Daniel K Inouye Solar Telescope will become operational providing spatial details of magnetism and thermodynamic evolution throughout the solar atmosphere at scales never before accessed. On the synoptic scale, CoSMO, the Coronal and Solar Magnetism Observatory, will have a 1.5 m refracting coronagraph at its centerpiece to measure the vector magnetic field of the corona above the limb, complemented by photospheric and chromospheric magnetic diagnostics, coronal emission line, and K-coronagraph measurements. The CoSMO suite was designed with operation and academic space weather forecast requirements in mind.

\item\textbf{What are the challenges in robust and routine acquisition and processing of relevant data?}

The most prevalently discussed challenge lies in the cross-referencing of the numerous observational datasets and observational databases. Quality of the latter is not assured and very much based on the diagnostics applied. As a result a high level of documentation, standardization and communication are required to advance common goals.

\item\textbf{What are the computational (real-time) bottlenecks of data analytics for forecasting?}

The group was mostly unaware of these, but suggested that the formatting/reformatting/regridding and re-averaging of data for comparison and assimilation into the simulations was a significant bottleneck.

\item\textbf{What are the challenges in quantifying data quality (e.g., calibration, uncertainty quantification, etc.)?}

Numerous items have been discussed above about the standardization of approach, documentation of the various calibrations steps, changes and algorithms. One approach commonly taken by space missions is that the Level-0 data are provided along with a calibration/correction algorithm that is updated and maintained in a community repository. In this way data can be reprocessed rapidly should improved calibration be released.

\item\textbf{What are the algorithmic developments that can enable ingestion into first-principles models?}

The group was mostly unaware of these.

\item\textbf{What is the potential of incorporating additional (new) data sources for forecasting?}

In the context of space weather forecasting, many things like chromospheric and coronal magnetograms and plasma environment data, such as those derived from inversions of observational data, have the potential to significantly improve the initial and evolving boundary conditions of the forecast model. However, in an MHD model, a methodology much be developed to assimilate these variables into the models on an appropriate timescale. That activity is non-trivial.

\item\textbf{What are the technical challenges in integrating remote-sensing and in-situ solar measurements?}

The integration of in-situ observations is a considerable challenge, indeed multi-sampled space is likely required to make significant advances although alternate approaches can be used. For instance, it will be a considerable challenge to accurately forecast the environment through which the Parker Solar Probe and Solar Orbiter will fly, but their data is of such novelty and value that great effort will be made. Much will be learned in the adaptation of the models to get close to matching the observations made by those spacecraft. Increasing model resolution upward to match that of the data is needed for more meaningful comparison to the observations.
\item\textbf{What is the role of emerging advances in data science such as machine learning and feature recognition in producing products that can be incorporate into first-principles models?}

The group was unclear about this question. However, it was declared that information from machine learning could inform physical models. Machine learning can track a vast number of parameters that humans can’t keep in mind and seek correlations and correspondences that we may not have considered. Deep learning parameters can be referred back to the observational inputs, so that a machine learning model fed by a physics-based model could inform the input, in principle. It was accepted that feature recognition is a more repeatable way of getting inputs into models.

\item\textbf{What future research directions have the highest potential for accelerated progress toward improved data products for incorporation into physical models?}

Many activities can accelerate progress in space weather. First and foremost is likely the identification of processes responsible for the population and acceleration of the solar wind, since it is the background state of the entire system. Further, improve the use of existing photospheric vector magnetograms by including for chromospheric vector magnetograms and merging with coronal emission and spectropolarimetric measurements. Providing synoptic maps of filaments magnetic fields (strength and chirality) removes the need to inject models with ad-hoc flux ropes. Algorithms for such activities are computationally expensive, but machine learning and pattern recognition methods could be brought to bear to accelerate progress.

Observations to advance understanding and (eventually) forecast skill/ accuracy include, but are not limited to: heliospheric imagers for region between Sun and 1 AU, and especially a polar point of view of the system; EOVSA/FASR radio data with sensitivity to coronal magnetic field, high-energy particles; IPS arrays for heliospheric measurements; Metric radio imaging observations of type II and type III bursts; ALMA observations for chromospheric conditions, input to source of solar wind; CoSMO measurements of the coronal vector magnetic field, spectroscopic plasma properties, and coronal wave environment for wind studies.
\end{itemize}

\section{Computer Science Group Breakout Session Report}
The data analytics breakout group consisted of researchers with terminal degrees in computer science, computer engineering, electrical engineering, physics and astronomy. The members of the group specialized in data base systems, information retrieval, big data, distributed computing, data science, computer vision and data visualization. During the workshop, these scientists participated in the discussions with two other groups, but also worked in separate settings as data analytics group to provide their input into this white paper. The following sections contain an overview of the discussions conducted by this group, with a list of important research challenges ranked based on the priorities defined by this particular group. All recommendations presented below have been endorsed by the entire data analytics breakout group.

\subsection{The Highest Priority Need – Creation of Data Set Benchmarks}
Benchmark datasets are an integral part of the machine learning and data science (ML/DSc) research. Now, there is an abundance of publicly accessible benchmarks for different tasks involving big data analytics (such as computer vision \citep[FERET,][]{FERET}, \citep[CIFAR,][]{CIFAR}, or time series data mining \citep[UCR,][]{UCR}). The main value of benchmark datasets comes from the fact that they provide a steady foundation for quantitative and repetitive comparisons of otherwise often highly varied machine learning (ML) approaches to space weather prediction. Making such “frozen in time” benchmark data sets broadly available to research communities will enable us to perform reproducible, comparative analyses of different methods across various studies, and help to build standardized and fair comparison mechanisms between otherwise different approaches. Moreover, it will help with the broadening an engagement computer and data scientists (DSc), as these data benchmarks will provide a solidfoundation for dialog between solar and heliospheric communities and many interested ML/DSc researchers interested in interdisciplinary science but without immediate access to space weather experts. We have in mind a large number of ML/DSc researchers, who may even be able to access space weather data repositories, but may not want to perform the hefty pre-processing tasks required to transform observational data to formats and types used by the ML/DSc communities, while the data benchmarks from other domains are easily available in more familiar to them formats. Therefore benchmark data should be provided in formats and types that are easily used by the ML/DSc communities, lowering the data usage impedance, making communication between the ML/DM and solar and heliospheric communities easier, and promoting inclusion of reproducible and comparative studies when new approaches to space weather prediction are being proposed. The data scientists' interest in easily-accessible benchmark datasets should never be confused with the advocacy for converting the vast repositories of solar data from one format to another. The intention is neither to modify data formats used by operational data providers nor to propose changes to the way solar physicists process their data,  Instead the panel focused on new ways in which this community can share the subsets of solar data to encourage and develop novel and fruitful collaborations with data scientists.
One important aspect of data benchmarks is that these datasets are usually created for a particular set of machine learning tasks. For example, MNIST dataset \citep{MNIST} is a large database of handwritten digits frequently used for training Deep Neural Networks (DNNs). Thus, one focal point of benchmarking efforts should be the task we intend them to be used for. Additionally, a common metric or set of metrics could be defined in order to easily and quantitatively compare multiple approaches.
Before presenting our discussions, we need to applaud solar/heliospheric physicists for their recent efforts put into deriving benchmark datasets from solar repositories. Two great examples are (1) the work of Monica Bobra (available as a Jupyter workbook) on flare and CME data sets, as well as (2) the SDO synoptic set of immutable 1k x 1k EUV images, publically available at  \url{http://jsoc2.stanford.edu/data/aia/synoptic/}.
Considering multiple aspects of our discussions, we would recommend taking the following points under consideration, when creating the benchmark datasets for comparative space weather data analyses:

\begin{itemize}
\item Benchmarking efforts should focus first on specifying the clear definitions of events (as mentioned in call for establishing benchmarks for space weather events on page 5 of the Space Weather National Strategy report, \url{https://bit.ly/2pxDnIp)} that may have impact on our infrastructure, and are of significant interest to our communities. These include, but are not limited to, solar flares, solar energetic particles (SEPs), and coronal mass ejections (CMEs), with event magnitudes and associated recurrence intervals.

\item Benchmark datasets should be created with: (1) the intention of targeting a particular ML/DSc community through the data types these dataset include (e.g., to attract computer vision experts to work on automatic solar detection models, big data sets are needed with solar events precisely marked as ROIs (regions of interest) on the solar images. Such data sets can be currently generated by combining outputs of computer vision modules run in SDO pipeline with original images from SDO AIA cameras, as demonstrated in \citep[LSDO,][]{LSDO}. (2) the usage of data dissemination methodologies these communities prefer to use (e.g., the vast majority of time series data miners uses UCR archive for comparative studies \citep{UCR} and maybe even (3) specific data analysis tasks in mind (e.g., information retrieval, classification, regression, clustering).
\item These data benchmarks may contain time series data (or non-evenly spaced sequences), images, or series of images (video), and/or be represented in the format of spatial (or spatio-temporal) raster data (matching so called \textit{field} data models) or vectors (\textit{object} data models).
\item Another interesting option for data benchmarks generation is combining a diverse  set of relevant data sources (e.g., from multiple space and ground observatories) for a heterogeneous benchmark big dataset containing multiple types of solar events. We thought this was a very interesting idea, although some participants expressed a valid concern that creating such a massive dataset may take some significant effort and computing power and may require the maintenance of a large data repository, while in practice the use of such big data set by the ML/DSc community could end up being quite limited due to the overwhelming size, and linked to it the costs of storage and data transfers. In such case, we thought that limiting the scope of such heterogeneous datasets temporally can be a good alternative (e.g., preparing the slices of such heterogeneous data for a particular time intervals, matching lifespans of particular solar events that are of significant interest to the broader space weather community). However, this poses another research challenge - figuring out which time window(s) would be the most useful for such a benchmark dataset.
\item Final point worthy of a significant consideration when developing benchmark datasets is the nature of the task which the dataset is supposedly aiming to help solar and heliospheric communities. From our discussions, we concluded that purely scientific explorative studies (research) has most interest in the highest quality data without the strong priority given to the data being available in a real-time (i.e. immediately when the events happen), while the more practical predictive analyses (intended for operational forecasting) may purposely focus on datasets that can be used for creating the most reliable models, and are (at the same time) coming from observational instruments that can be maintained for a longer time, and can successfully address the practical challenges related to near real-time data processing (i.e., immediate online data availability, permanently reliable data sources).

\end{itemize}

\subsection{Taking Advantage of State-of-the-Art Machine Learning Approaches}

Although data scientists can sometimes observe occurrences of a trade-off between attaining the highest accuracy (defined as the level of agreement between the prediction generated by ML model(s) and the truth, represented by actual observations) possible from machine learning methods and attaining a human-interpretable result from the same group of data mining methods, we believe that both the "white box" as well as the "black box" machine learning models can serve well our space weather community.  While it appeared that the operational community may be more interested in the accuracy and the reliability (defined as the average agreement between the predicted values and the observations over time) at the potential expense of interpretability of the machine-derived prediction processes, it also appeared that the research community may be more interested in interpretability of data-derived decision making process, due to potential for interpretable machine learning (i.e., "white boxes") to stimulate new science questions and, in consequence, potentially improve human understanding of the space weather.  We want to make sure it is clearly understood that this discussion was conducted in reference to the accuracy/reliability of the outputs of machine learning methods (as defined by probability of correct classification or similar metrics), and not the accuracy of the input data (as defined by fidelity of the processed data products made available by our data providers), which was beyond the scope of our discussions.  While we all should ideally strive for the highest possible accuracy and reliability while maintaining human interpretability of ML models, it was agreed it is worth to consider solutions across the broad spectrum of machine learning methods, that is between uninterpretable, but often highly accurate, methods on the one end and interpretable, but sometimes less accurate, methods on the other.  It is worth noting that human interpretable machine learning algorithms may necessarily be simpler formulations (e.g., decision trees) than the highest accuracy algorithms (e.g., deep neural networks) and that those simpler formulations may sometimes be not capable of capturing complexities of the underlying data.
While deep neural networks have recently produced state-of-the-art results in many complex classification and regression tasks, our understanding of their actual decision making process (i.e., of the computations they perform on particular data while learning/training and then applying the learned knowledge/testing on the data never seen before) is still somehow limited. A particularly poorly understood issue deals with the intermediate knowledge representations they build and their relationship with observable variables. Research is needed to analyze these complex representations in order to extract physical insight about the phenomena they model.
At the other end of the spectrum, models such as decision trees and their extensions produce results that are easily interpretable, albeit often not as accurate as those of deep models (especially for highly complex data). Thus a complementary research direction is recommended, to develop (on one side) new ways of enhancing the accuracy of interpretable models in a way that allows them to yield competitive prediction results, and to increase (on the other side) our understanding of complex machine learning models, such as deep neural networks’ structures, after they have been trained on complex space weather data.
With the astonishing successes of deep neural networks (DNNs) in the recent decade in the areas of computer vision and natural language processing, many participants pointed out their interests in investigating DNNs’ applicability to space weather predictions. Deep neural networks are currently known to be able to produce very accurate results in complex classification and regression tasks by modeling the complex relationships between inputs and outputs in a purely data-driven fashion. They often need very large amounts of training data in order to produce these highly-impressive results, and significant computational power is needed to build (i.e. train) accurate models. Moreover, in many cases, there can be knowledge available about the physical phenomena the DNNs are expected to model that is not appropriately exploited by these algorithms. Research is needed to develop ways to incorporate this knowledge into the deep learning processes, to reduce the need for training data and expensive computational resources, and to ensure that DNN-based prediction model perform well when presented with rare events for which we may have no sufficient data. One highly interesting and extremely ambitious research challenge was thrown down during our discussions. That is to implement a DNN model that could learn from \textit{all} the data space weather community have collected through decades, and what would it take to make it actually work. Many of points discussed in this white paper (e.g.: (1) heterogeneous big data benchmarks discussed in Sec. 6.1, (2) distributed machine learning algorithms, which are mentioned in Sec. 6.2.1, and (3) creation of centralized data center, mentioned in Sec.6.7.1) directly relate to this newly proposed \textit{Big Space Weather Deep Neural Network} challenge.

\subsubsection {Scalable and Distributed Machine Learning Algorithms for Space Weather Big Data}

With the enormous growth of solar data, initiated by the SDO observatory, and guaranteed to continue due to currently build other observatories, ongoing research on distributed machine learning may offer a valuable approach for solar big data analysis. In recent years, many practices have demonstrated a trend that more training data and bigger models tend to generate better accuracies in various data mining applications. However, it remains a challenge for the vast majority of machine learning researchers and practitioners to manage and train their machine learning models from huge amounts of  heterogeneous data. The task usually requires a larger amount of computational resources (data storage, RAM/volatile memory, parallel computing capabilities, fault tolerance, etc.) than can be offered by a single machine. In addition, we would like this setup to scale well with increasing volume of training data or model complexity.  It was pointed out by the panel that the anlysis of Solar Big Data may benefit from distributed machine learning algorithms, tools, resources and platforms (e.g., Apache Spark cluster) and state-of-the-art research that is being conducted in this direction.  Computer scientist have already developed some libraries that operate on top of distributed environments such as Apache Hadoop and HDFS or Apache Spark for an improved in-memory computations, but much more work is needed. Good example is the library SparkML, which allows big data analysts to focus on tuning algorithms, rather than programming them in a distributed manner. We thought that it may be beneficial for our space weather collegues to take a look at some of the above-mentioned technologies. The need for the state-of-the-art cyberinfrastructure is made quite evident when machine learning models need to be trained and deployed at the Space Weather Big Data scale.

\subsubsection {Challenges with Employing Machine Learning Models in Operational Settings}

While it is often desirable and convenient to have automated or parameter-free algorithms that require zero or minimal human intervention, in certain applications such as space weather forecasting, it may be beneficial to keep human experts involved. This can be achieved through several aspects. Firstly, allowing models or algorithms to incorporate user input (e.g., known constraints coming from our knowledge of physics behind solar events, human-suggested thresholds for pattern mining algorithms) would ensure that the data mining algorithms adhere to the physics behind the space weather phenomena. Secondly, allowing scientists to provide feedback on the outputs of the algorithms (e.g., the resulting model parameters, usability of discovered patterns) can help validate the results and re-confirm or deny (through large scale statistical studies) commonly believed relationships between space weather events, and/or fine-tune the algorithms to achieve a better accuracy and reliability. Finally, when data on certain event instances are rare or prohibitively expensive to obtain, active learning and transfer learning can be used to help mitigate these issues.

\subsubsection {Event-guided Image Data Compression}

Event-guided data compression is another interesting concept that may allow for a more efficient data storage, transmission, retrieval, and (in consequence) more effective machine learning from large-scale solar data. Traditional data compression techniques (such as JPEG) are not exploiting the redundancies existing in large scientific data sets. In contrast, recently researched unsupervised machine learning methods, such as deep encoders (e.g., \citet{7906393}) may allow for automatic discovery of longer-term correlations occurring in the large-scale image data repository which could potentially lead to a much higher data compression than currently offered.
On the other hand, supervised machine learning methods can be used to identify regions of interest (ROIs) in the solar data, which are critical for modeling and prediction in the later stages. Lossless compression can be used for these ROIs while much higher, lossy compression could be used to compress other regions of the Sun, which may be of lesser importance. In result, a much higher overall compression can be achieved over the entire dataset than can be achieved by currently popular data compression algorithms. Direct processing and learning from these compressed data could eliminate the need to decompress the data which would help to reduce delays due to data transformations as well as the storage requirements.

\subsection{Relevance of Space Weather Big Data to Different Machine Learning Communities}
\subsubsection{Computer Vision Community}
The open availability of the massive NASA image datasets makes them highly attractive for computer vision community. However, the initial high cost (man hours) required to collect and curate the data is the biggest barrier for computer vision scientists interested in solar and heliospheric data mining field to work on such rich datasets. By having curated data benchmarks, as outlined in Sec. 6.1, this obstacle can be removed and we would be able to attract top researchers doing state-of-the-art work to participate in the efforts of our community.
In 2013 there has been a first effort to release datasets \citep{6738896}, but it was not until \citet{LSDO} that a large enough dataset became easily accessible to the computer vision community. With the release of other dataset benchmarks, currated by solar domain experts, we anticipate a larger amount of synergy between these two fields. The benchmark datasets will facilitate the work of computer vision experts, and more and more computer vision results can be achieved that are desired by solar and heliophysics domain experts (e.g., accurate and large-scale identification of multiple types of solar events on the Sun).

\subsubsection{Time Series Data Mining Community}
The discovery of novel, previously unknown patterns from time series data has received a great amount of attention from researchers in the data mining community in the past two decades. Finding novel patterns, regularities, or rules from data can provide insights to the decision maker, help us to know what to expect for future events, and allow more accurate and targeted space weather monitoring. One example of such novel group of patterns in time series data are \textit{motifs}, which are frequently occurring, highly interpretable subsequences. They can be used to build better classifiers, identify other interesting patterns such as anomalies and sequential rules for event predictions. Another example is time series \textit{shapelets} — subsequences that are “maximally representative” of a class. Shapelets generalize the lazy nearest neighbor classifier to an eager, decision-tree-based classifier which can improve the classification speed and interpretability of the data-derived space weather prediction process. In general, it has been shown in other domains that classifiers capable of learning local, contextual patterns in time series data can provide solutions that are more accurate, robust, and interpretable, when compared to traditional approaches (e.g., Fourier trasnform, or moving avergaes). We believe that these (never applied on solar data) techniques from time series data mining may have a huge impact on space weather forecasting or classification.

\subsubsection{Spatio-Temporal Data Mining Community}
The management and mining of moving objects have been extensively studied in computer science and GIS literature for a wide range of applications \citep{Guting, Saltenis, Lee, Shekhar}.

A potentially impactful research avenue for solar physics and data science communities is generation, management and mining of spatio-temporal solar event metadata. The solar event metadata (vector data) contain temporal and spatial information for various solar events, as well as the time of occurrence and locations of the events. The location information for many important event types, such as active regions, coronal holes, or filaments, is rather detailed and is in the form of spatial polygons created from location encoding chain codes. Recently, a data-driven tracking module \citep{KEMPTON} and spatio-temporal interpolation techniques \citep{Filali} have been developed to connect the solar event records belonging to the same phenomena and to enhance the quality of location data. With these recent developments, we can now model the solar events as moving region objects whose spatial locations continuously evolve over time.
Using these moving regions of solar events, new spatio-temporal frequent pattern types and mining techniques are created specifically for solar events \citep{Aydin, AYDIN2015136}. Another research direction, specifically for retrieval tasks, is the augmentation of spatio-temporal data to continuous streams of solar image data and utilizing the video retrieval techniques\citep{Hu}.

We anticipate more studies will be conducted on spatio-temporal aspect of solar event data with the greater availability of spatio-temporally enhanced solar event metadata. We believe that the spatio-temporal data mining from solar events may help our colleagues from solar and heliospheric disciplines to better understand the relationships among different solar event types and how the evolutions of particular solar events may affect the nearby events.

\subsection{Infrastructure for Machine Learning from Solar Big Data}
 As a complicated game of trade-offs, choosing between High Performance Computing (HPC) versus Cloud Computing, and between using Computing Clusters and Graphics Processing Units (GPUs) is not an easy task. Proper choice always comes down to the particular application and its specific requirements. The costs of any of the above mentioned solutions can go quite high if the data processing methodology is not correctly paired with an undertaken task, and the computational requirements of used machine learning algorithms are accurately evaluated. However, when properly optimized, a perfect solution with appropriate cost/value can be usually found. In this section, we will briefly discuss a few of these computing infrastructures, pointing out elements that may be of importance for researchers interested in Machine Learning from Solar Big Data.

\subsubsection{Centers for Holding and Processing the Solar Big Data}
As solar datasets become larger, it becomes increasingly difficult to download data for local processing.  Moreover, certain modern machine learning algorithms require large amounts of data and computational resources to train them. This may generate the need to have substantial computing power devoted to data analytics in proximity to where the big data is stored. Further, the challenges of predicting space weather may require multiple heterogeneous datasets gathered all along the Sun-Earth line.  A centralized location that holds these data, would make them easily available for analysis to researchers in all communities, and even enable machine learning experts to freely run their data-driven analysis at its equipment. This setup could highly benefit the space weather community.
The easily accessibility to computing and storage servers with direct and fast access to the original data will greatly enhance the data processing community’s ability and efficiency in analyzing solar big data.  Many state-of-the-art algorithms are data hungry, meaning large subsets of data need to be processed, indicating an advantage to co-locating the computing facilities with the data storage servers.  Furthermore, many state-of-the-art algorithms are iterative in nature and amenable to parallelization and can greatly benefit from processing on multi-core and/or multi-GPU machines, indicating a need for significant computational power within those computing facilities.  Lastly, many of the machine learning algorithms may generate intermediate results that may be of interest to other members of our community, requiring significant local storage for FP models' outputs (separate from the original data repository) and close to the computing infrastructure.

\subsubsection{HPC, Cloud Computing, Computing Clusters and GPUs}
Cloud computing allows for offload of the heavy processing of training machine learning algorithms to the server, which might be equipped with GPUs and computing clusters. This way, mobile and distributed computing become possible on “thin” clients.
The same can be said about computing clusters, as the processing and storing of data is distributed among the multiple nodes and jobs are scheduled by “thin” clients. However, the cost factors always depend on the application. In-house local clusters require a substantial initial investment (already made by many research institutions) and the hardware becomes obsolete very quickly, versus cloud services that require very low entry costs (you pay what you use), are great for some preliminary studies, and the underlying hardware is frequently replaced by the service provider, at their expense. Another considerable factor is the management of such infrastructure, a local cluster needs an in-house person to manage it and make sure it operates at peak performance, versus cloud infrastructure that is managed by the provider at no extra expense to the local institution.
In terms of the argument of traditional CPU-based computing (cluster or non-cluster) versus GPU-based computing, it also depends on the size and type of a problem. For something that is purely computationally based, a GPU has a clear advantage over traditional CPU computing due to the nature of GPUs, however, when large volumes of data need to be loaded in memory, the GPUs have bottlenecks as their memory allocations are considerably smaller than in the CPU-based architectures. The good news are that, at least for cloud environments, these trade-offs can be relatively easily tested before buying any expensive hardware and can be scaled down or up based on the actual needs.
The major downside of cloud-based solutions, however, is the storage costs associated, and while all the compute nodes and settings can be “turned-off’, there will always be storage costs associated with having the data available there. The storage costs make often Cloud Computing cost high for both big data machine learning as well as modeling communities. While the first community had to deal with high costs of uploading the big data to perform data mining in the cloud and then just need to download small (storage-wise) data-driven models, the modeling community uploads relatively small (storage-wise) code, but need to download massive outputs generated by their models.
Overall, the need for computational resources is evident and imperative, however, what kind of resources and how to better use them is something that needs to be very carefully analyzed in advance of each particular project.
\section*{Acknowledgements}
The workshop was funded through the NSF grant AGS-1812964 to the New Jersey Institute of Technology (NJIT).
\bibliography{rref}
\end{document}